\documentclass[aip,apl,twocolumn,superscriptaddress,floatfix,showpacs]{revtex4}
\usepackage{graphicx}
\usepackage{amsmath,amssymb}
\usepackage{color}

\begin{document}

\newcommand{\un}[1]{\,\mathrm{#1}}
\newcommand{\tp}{\ensuremath{t_\mathrm p}}
\newcommand{\tw}{\ensuremath{t_\mathrm w}}
\newcommand{\gl}{\ensuremath{\Gamma_\mathrm L}}
\newcommand{\gr}{\ensuremath{\Gamma_\mathrm R}}
\newcommand{\gi}{\ensuremath{\Gamma_1}}
\newcommand{\gii}{\ensuremath{\Gamma_2}}
\newcommand{\dd}{\ensuremath{\mathrm d}}
\newcommand{\e}{\ensuremath{\mathrm e}}
\newcommand{\comment}[1]{\textcolor{green}{#1}}

\title{Radio frequency pulsed-gate charge spectroscopy on coupled quantum dots}

\author{D.\ Harbusch}

\affiliation{Center for NanoScience and Fakult\"at f\"ur Physik, Ludwig-Maximilians-Universit\"at M\"unchen, Geschwister-Scholl-Platz 1, D-80539 M\"unchen, Germany}

\author{S.\ Manus}

\affiliation{Center for NanoScience and Fakult\"at f\"ur Physik, Ludwig-Maximilians-Universit\"at M\"unchen, Geschwister-Scholl-Platz 1, D-80539 M\"unchen, Germany}
\author{H.\,P.\ Tranitz}

\affiliation{Institut f\"ur Experimentelle Physik, Universit\"at Regensburg, D-93040 Regensburg, Germany}

\author{W.\ Wegscheider}

\affiliation{Laboratory for Solid State Physics, ETH Z\"urich, CH-8093 Z\"urich, Switzerland}

\author{S.\ Ludwig}

\affiliation{Center for NanoScience and Fakult\"at f\"ur Physik, Ludwig-Maximilians-Universit\"at M\"unchen, Geschwister-Scholl-Platz 1, D-80539 M\"unchen, Germany}

\begin{abstract}
Time-resolved electron dynamics in coupled quantum dots is directly observed by a pulsed-gate technique. While individual gate voltages are modulated with periodic pulse trains, average charge occupations are measured with a nearby quantum point contact as detector. A key component of our setup is a sample holder optimized for broadband radio frequency applications. Our setup can detect displacements of single electrons on time scales well below a nanosecond. Tunneling rates through individual barriers and relaxation times are obtained by using a rate equation model. We demonstrate the full characterization of a tunable double quantum dot using this technique, which could also be used for coherent charge qubit control.

\end{abstract}

\maketitle

\section{Introduction}

Solid state quantum information processing based on single electrons in coupled quantum dots (QDs) demands very precise control of the quantum states involved \cite{Nielsen2000,Macchiavello2000}. On the one hand, it is necessary to perform many qubit operations within the coherence time. On the other hand, to address quickly specific qubits in a large array and detect their response poses a serious experimental challenge. On the detection side, single-shot readout of charge states of QDs has been demonstrated up to a bandwidth of 10\,MHz which is limited by the detector noise \cite{Clerk2003,Elzerman2004,Vandersypen2004,Gustavsson2006,Vink2007,Reilly2007,Cassidy2007,Kueng2009,Barthel2010}. Shorter time scales can be resolved using a time-averaged detection while an excitation pulse cycle is continuously repeated \cite{Nakamura1999,Petta2005a,Gorman2005,Hayashi2003,Shinkai2009}. In this way, charge coherence times below 1\,ns have already been resolved in transport spectroscopy measurements \cite{Hayashi2003,Shinkai2009} where the chemical potential of one lead of a double QD was pulsed while the average current through the double QD was measured. However, this technique, based on pulsing ohmic contacts, will fail in more complex devices which consist of a larger number of coupled QDs with only few ohmic contacts (a more realistic circuit for quantum information processing). The accessibility of individual QDs, needed for the initialization or readout of specific qubits, requires more involved techniques such as high-bandwidth locally resolved charge spectroscopy. In this scenario, several close-by gates of a complex nanostructure can be individually pulsed while the conductance of a capacitively coupled quantum point contact (QPC) \cite{Field1993} or alternatively a detector QD is measured. A main obstacle for high-bandwidth charge spectroscopy is cross-talk between adjacent gates, that is, leakage of a voltage pulse originally applied to one specific local gate into other components of the nanostructure.

We developed a high-bandwidth pulsed-gate technique with square pulses much shorter than a nanosecond. Here, we use it to study time-resolved energy relaxation and quantum-mechanical tunneling of a single electron in a double QD. To avoid the disadvantages of current spectroscopy, we perform charge spectroscopy using a nearby QPC as a detector. A similar pulsed-gate technique has already been used for the control of spin qubits \cite{Petta2005,Johnson2005,Hanson2007}. However, compared to our measurements, the pulse durations and repetition periods reported there were much longer, hence, imposing smaller challenges to radio frequency (rf) components. In addition, these earlier and following works relied on a qubit initialization via charge exchange with the leads, which involves pulsing between three different charge configurations of the double QD. In contrast, we pulse between only two configurations and work at a constant overall charge. In this case, charge exchange with the leads is no longer required, and the tunnel coupling to the leads can be tuned to be very weak. We expect this effective decoupling from the bath of two-dimensional electrons to ultimately lead to longer coherence times. Here, we demonstrate a characterization of the double QD parameters relevant for the control of a charge qubit. We use the same pulsed-gate technique which will be employed to actually operate qubits, hoping to increase the overall efficiency of a future quantum computer. Introducing a rate equation model allows us to obtain from our data tunneling rates of individual barriers and energy relaxation times. Moreover, we are able to distinguish between possible charge relaxation channels, which can be tuned via gate voltages. This is especially important for future qubit applications where specific relaxation channels might be temporarily opened for initialization or readout, while long energy relaxation times are required during coherent manipulation.

\section{Nanodevice and setup}

Mapping the applied pulses with a short rise time ($\sim70\,$ps) to a time-dependent electrical potential at the double QD requires high rf bandwidth at cryogenic temperatures. Our gate geometry and rf sample holder are designed to minimize the detrimental cross-talk and optimize impedance matching. At the same time, using rf wiring in our dilution refrigerator, we have extended the bandwidth of our measurement to beyond 10 GHz (presently limited by the use of micro miniature coax connectors).

\begin{figure}
\includegraphics{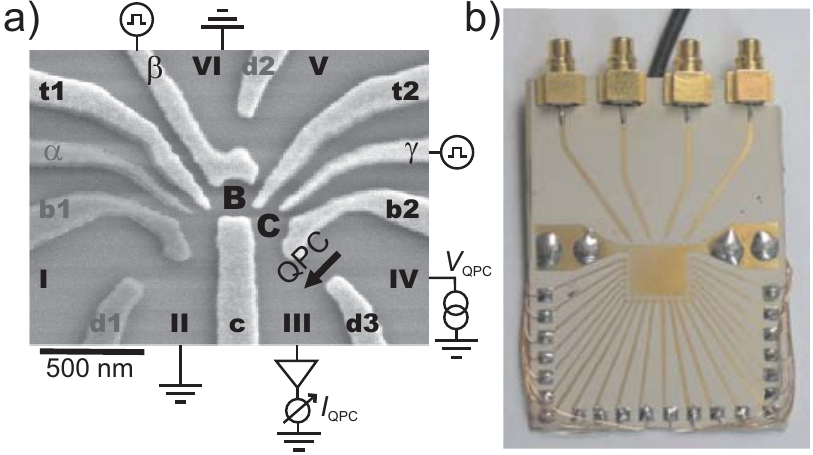}
\caption{(a) A scanning electron micrograph of a device nominally identical to the investigated one. Details in the main text. (b)  The radio frequency sample holder with four impedance-matched micro strip lines and 22 low bandwidth connections. The central gold square is electrically shorted to the copper plated backside of the circuit board. The sample would be mounted on the central gold square and wire bonded.}
\label{fig1}
\end{figure}

Figure \ref{fig1}a
shows a scanning electron micrograph of the surface of the double QD device based on a GaAs/AlGaAs heterostructure. It contains a two-dimensional electron system at a temperature of $T_\mathrm{el}\simeq 130\,$mK. The metallic gates can define electrostatically a few-electron triple QD \cite{Schroer2007,Gaudreau2009}. However, in the present experiment, we only define a double QD (labeled B and C) and implement one of three QPCs as a charge detector \cite{Field1993}. Therefore the three darkened gates ($\alpha, b1$ and $d1$) in Fig.\ \ref{fig1}a are connected to the electrical ground. All ohmic contacts (labeled  I -- VI) are also grounded with the exception of contact IV, which is biased with $V_\mathrm{QPC}\simeq-50\,\mu$V. The resulting dc current, $I_{\mathrm{QPC}}$, flows through the QPC to the grounded contact III , while current flow through the double QD is negligibly small even under resonance conditions. $I_{\mathrm{QPC}}$ depends on the charge state of the double QD. We reduce back-action of this charge sensor onto the double QD by limiting the detector sensitivity (at still high contrast) \cite{Harbusch2010}. In detail, we only apply a low voltage $V_\mathrm{QPC}$ and operate the QPC at a small conductance of $G_\mathrm{QPC}\ll 0.5\times 2e^2/h$.

Gates $t1$, $\beta$, $t2$ and $\gamma$ are connected to rf coaxial cables. The dc voltage applied to these gates can be modulated via high-bandwidth bias tees (Agilent 11612A) operated at room temperature. Pulse trains are produced with a pattern generator (Agilent 81134A) that provides a minimal rise time of $70\,$ps. A main technological challenge is the precise transmittance of the pulse shapes generated at room temperature to individual gates of the nanoscale device at low temperature. When choosing rf cabling, a compromise has to be made between a small thermal conductivity and a low rf damping. We use stainless steel coaxial cables with a silver-plated inner conductor from room temperature to $T\simeq4.2\,$K and superconducting coaxial cables at lower temperatures. For thermal anchoring of the inner conductor of the coaxial lines at various temperatures, we have developed impedance-matched gold strip lines on sapphire sheets with insertion losses below $0.4\,$dB. These devices are virtually resonance free up to 12 GHz. Our broadband sample holder \cite{comment-1} (Fig.\ \ref{fig1}b) contains four micro strip lines which are optimized -- with the help of numerical calculations using the commercial software \emph{SONNET} -- for transmission of short pulses and minimal cross-talk \cite{Reilly2007}. The sample is mounted on the central gold square and electrically connected via short bond wires.

\section{Stability diagram and pulse calibration}

\begin{figure}
\includegraphics{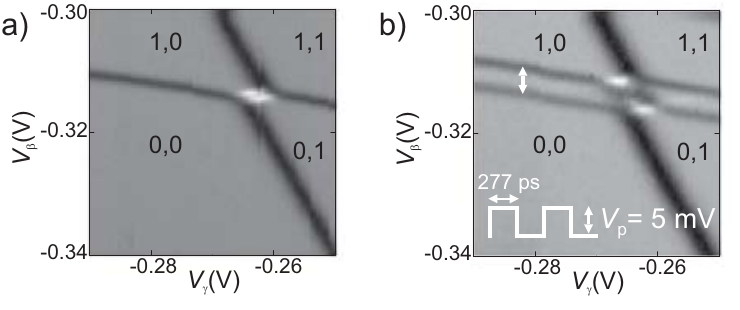}
\caption{Charge stability diagrams of the double QD. Plotted in gray scale as a function of gate voltages $V_{\beta}$ and $V_{\gamma}$ (see Fig.\ \ref{fig1}a) is the transconductance $\dd I_\mathrm{QPC}/\dd V_{\beta}$, which has been obtained by numerical differentiation of the dc current $I_\mathrm{QPC}$. $\dd I_\mathrm{QPC}/\dd V_{\beta}\simeq0$ at the gray background, $\dd I_\mathrm{QPC}/\dd V_{\beta}<0$ on darker lines and $\dd I_\mathrm{QPC}/\dd V_{\beta}>0$ at brighter regions. Pairs of numerals 0 or 1 depict the number of electrons charging the QDs B,C in the ground-state configuration. (a) No pulses applied. (b) $V_{\beta}$ is modulated with a symmetric continuous square wave signal, for which the waiting time \tw\ between two pulses equals the pulse duration \tp, at a period of $f_0^{-1}=\tw+\tp=554\,$ps and an amplitude of $V_\mathrm{p}=5\,$mV (see inset and main text).}
\label{fig2}
\end{figure}

Figure \ref{fig2}a
is a plot of a section of a charge stability diagram of our double QD, where it contains between zero and two electrons, as a function of gate voltages $V_\gamma$ and $V_\beta$. The transconductance $\dd I_\mathrm{QPC} / \dd V_\beta$ (gray scale), obtained by numerical differentiation, clearly shows four charging lines (dark) separating stable ground-state configurations with zero or one electron in QD B and C, where pairs of numerals 0 or 1 depict the number of electrons charging the QDs B and C. The charge reconfiguration line (white) which separates configurations 1,0 and 0,1 appears as a double line because of external noise or back-action from the biased QPC \cite{Harbusch2010}. The measurement is repeated in Fig.\ \ref{fig2}b while gate $\beta$ (y axis) is modulated with a continuous square pulse train  at a repetition frequency of $f_0=1.805\,$GHz and an amplitude of $V_\mathrm{p}=5\,$mV. In this measurement, the pulse duration, $\tp$, equals the waiting time between two pulses $\tw=\tp=277\,$ps (see inset of Fig.\ \ref{fig2}b), where $\tw+\tp=f_0^{-1}$. We call this a symmetric pulse train in contrast to an asymmetric pulse train with $\tw\ne\tp$.

As a result of the symmetric pulse train, the stability diagram in Fig.\ \ref{fig2}b splits into two copies shifted along the y axis by the effective pulse amplitude $V_\mathrm p\simeq5\,$mV. This steady-state result is predicted (for $\tp=\tw$) by thermodynamics, even though our repetition rate is much shorter than the relevant relaxation rates -- in contrast to earlier pulsed-gate measurements \cite{Petta2005}. Using a calibration via nonlinear transport measurements, $V_\mathrm p$ can be converted to an energy shift of the electronic spectrum of the QDs which is $\Delta E\simeq 625\,\mu$eV. This value is large compared to the energy scale $h f_0\simeq 7.5\,\mu$eV corresponding to the pulse train repetition frequency. For the opposite case of $h f_0 > \Delta E$, effects like coherent destruction of tunneling would be expected \cite{Kohler2005}. For the limit of $h f_0 \gg \Delta E$ photon-assisted tunneling \cite{Wiel2002} would influence the outcome of similar measurements.

Comparison of the effective pulse amplitude $V_\mathrm p\simeq 5\,$mV with the actually applied modulation amplitude $\Delta V=6\,$mV quantifies the losses during the transmission of the pulses from the instrument at room temperature to the gate. The corresponding power attenuation of $20 \log(V_\mathrm p / \Delta V)\,\mathrm{dB}\simeq-1.6\,$dB can be traced back to the damping of the coaxial cables.

\section{Radio frequency sample holder }

\begin{figure}
\includegraphics{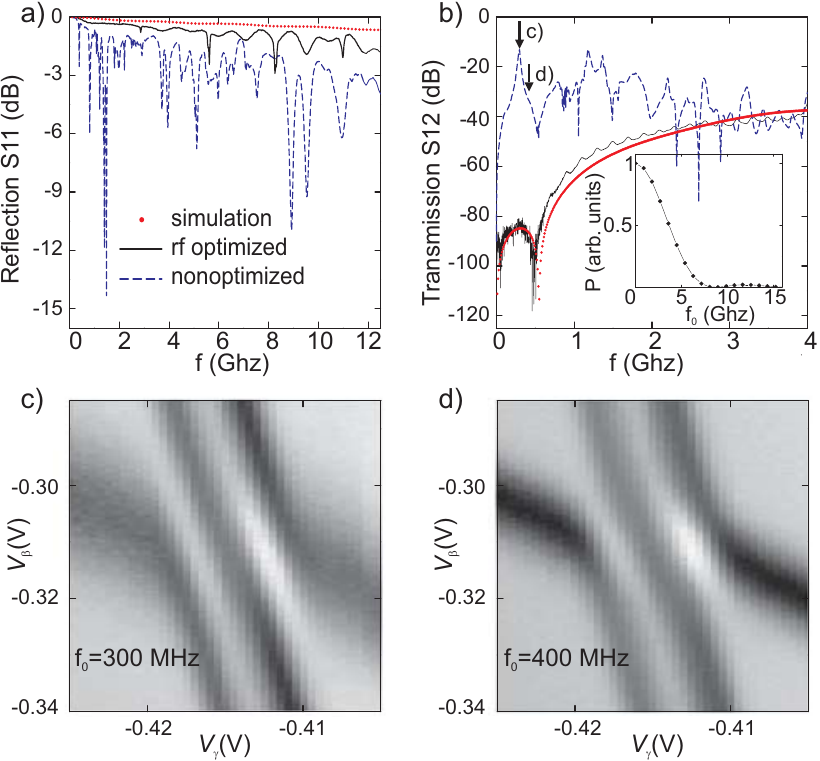}
\caption{(a) Reflection coefficient (S11 in dB) of a single open-ended lead of an rf sample holder (black solid line), optimized for minimal cross-talk and impedance matching, compared to that of a nonoptimized sample holder (blue dashed line) as a function of frequency. The (red) closed circles are a numerical simulation of the rf sample holder using the commercial software \emph{SONNET}. (b) Transmission (S12 in dB) through the two sample holders as a function of frequency. The transmission is measured between two adjacent but electrically isolated gates. The inset shows the calculated Fourier transform of a pulse train with $\tp=\tw=277\,$ps as a function of frequency, where $P$ is the power in arbitrary units. (c) A section of the double QD stability diagram, as in Fig.\ \ref{fig2}b, but using a nonoptimized sample holder while a symmetric pulse train with a repetition frequency of $f_0=300\,$MHz is applied at gate $\gamma$. (d) Measurement as in c) but at a slightly larger repetition frequency of $f_0=400\,$MHz (nonoptimized sample holder). The two frequencies are marked in b) by vertical arrows. }
\label{fig3}
\end{figure}

To elucidate the importance of an rf sample holder, in this section we compare its rf properties with a nonoptimized sample holder combined with a standard chip carrier. Pulsed-gate spectroscopy with the nonoptimized setup suffers from reflections and strong cross-talk between adjacent leads. This is illustrated in Fig.\ \ref{fig3}a
which compares network analyzer measurements of a nonoptimized sample holder (blue dashed line) and a rf sample holder (black solid line) engineered for optimal impedance matching and minimal cross-talk. Plotted is the logarithm of the power, reflected from a single open-ended lead (for the rf sample holder shown in Fig.\ \ref{fig1}b, on of the four strip lines at the top), in the frequency range $0<f<12\,$GHz. A perfect open-ended lead would show full reflection (0 dB), which is almost reached in a numerical calculation with \emph{SONNET} for our rf sample holder (red filled circles). Our rf setup is not perfect but its reflection losses are already much smaller than for the nonoptimized device. The strong oscillations of the reflection coefficient are caused by impedance mismatch resulting in resonances, phase shifts and radiation losses. The inset of Fig.\ \ref{fig3}b shows the calculated power spectrum of a typical pulse train ($\tp=\tw=277\,$ps, rise time $70\,$ps). It suggests that most power is contained in the spectral range $0<f<5\,$GHz, where the measured reflection properties of our rf sample holder (solid black line in Fig.\ \ref{fig3}a) are almost perfect. The imperfections at higher frequencies can nevertheless influence the pulse shape. For instance, strong cross-talk between the control gates can lead to even more serious pulse deformation. This cross-talk can be quantified by measuring the transmission between adjacent but electrically not interconnected leads. Figure \ref{fig3}b shows typical transmission curves for the two sample holders in the relevant frequency range. The measured cross-talk of the rf sample holder is fairly small and almost coincides with the numerical calculation done with \emph{SONNET} (red circles). The nonoptimized sample holder (blue dashed line) shows a much larger transmitted signal --- especially in the frequency range of the largest power density (see inset) --- and, in addition, strong oscillations.

Figures \ref{fig3}c and \ref{fig3}d demonstrate how the transmission influences the cross-talk in pulsed-gate experiments. The two plots show the same stability diagram section of our double QD measured with the nonoptimized sample holder under application of a symmetric pulse train to gate $\gamma$ ($x$ axis). While the nominal pulse shape with $\Delta V=4\,$mV and a rise time of 70\,ps is identical for both cases, the repetition frequencies are slightly different, namely $f_0=300\,$MHz for Fig.\ \ref{fig3}c and $f_0=400\,$MHz for Fig.\ \ref{fig3}d. As explained above (compare Fig.\ \ref{fig2}b), application of the pulse train results in a splitting of the more vertical charging line while the more horizontal one is only broadened. The broadening is, however, larger for the slightly smaller repetition frequency (Fig.\ \ref{fig3}c). The difference between the measurements in Figures\ \ref{fig3}c and \ref{fig3}d can be traced back to the transmission spectrum, which displays stronger cross-talk for $f_0 =300\,$MHz (arrows in Fig.\ \ref{fig3}b). Compared to measurements using an rf sample holder (Fig.\ \ref{fig2}b), the usage of the nonoptimized setup causes all charging lines to be broadened, which suggests imperfect pulse shape along with strong cross-talk. In summary, for quantitative charge spectroscopy using pulsed-gate measurements with subnanosecond time scales, the usage of a sample holder optimized for rf frequencies is essential.

\section{Pulse sequence}

\begin{figure}
\includegraphics{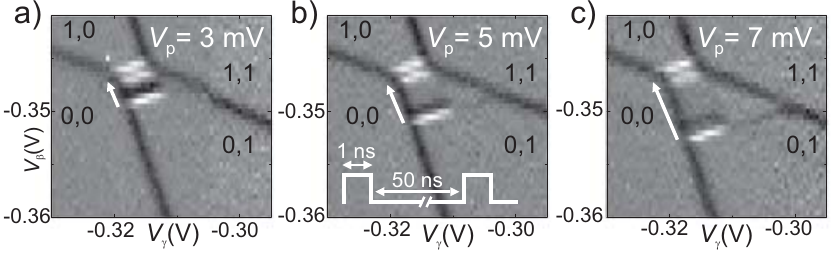}
\caption{Charge stability diagrams of the double QD as in Fig.\ \ref{fig2} but under application of an asymmetric pulse train ($\tw\gg\tp$). $V_{\beta}$ and $V_{\gamma}$ are both modulated in opposite directions. A square pulse of duration $\tp=1\,$ns is periodically applied after the waiting time of $\tw=50\,$ns. Effective amplitude and direction of the pulses are indicated by white arrows. The pulse amplitude $V_\mathrm{p}$ is increased from a) to c).}
\label{fig4_a}
\end{figure}

Figures \ref{fig4_a}a -- \ref{fig4_a}c
show measurements of the same region of the stability diagram as in Fig.\ \ref{fig2} but now for asymmetric pulse trains with $\tp\ll\tw$ ($\tp=1\,$ns and $\tw=50\,$ns). The pulses are applied in the direction of the white arrows parallel to the charging line of QD C. For this purpose, the two plunger gates $\beta$ and $\gamma$ are modulated simultaneously and with opposite sign ($V_\beta$ is increased while $V_\gamma$ is decreased) \cite{comment-2}. From Figs.\ \ref{fig4_a}a - \ref{fig4_a}c the pulse amplitude $V_{\un{p}}$ is stepwise increased. For the applied asymmetric pulse train, a naive thermodynamic model would predict the average charge occupation to almost correspond to the unpulsed ground-state configuration --- according to the local potential present between pulses. Within this model, we would expect to observe the same stability diagram as the one in Fig.\ \ref{fig2}a (with no pulses applied). This is indeed the case in Figs.\ \ref{fig4_a}a -- \ref{fig4_a}c, with the exception of an additional line (black-white) which corresponds to an average nonequilibrium charge configuration. The black-white line copies the charge reconfiguration line (white double line) in terms of having the same length and being parallel to each other. The distance between the black-white line and the charge reconfiguration line equals the effective pulse amplitude $V_\mathrm p$. Hence, the local potential \emph{during} the short pulses is identical to that on the charge reconfiguration line (in a measurement without pulses). Note that the black-white line is observed in transconductance and corresponds to a local maximum in the detector current $I_{\mathrm{QPC}}$ (see also Figs.\ \ref{fig6}a and \ref{fig6}b).

\begin{figure}
\includegraphics{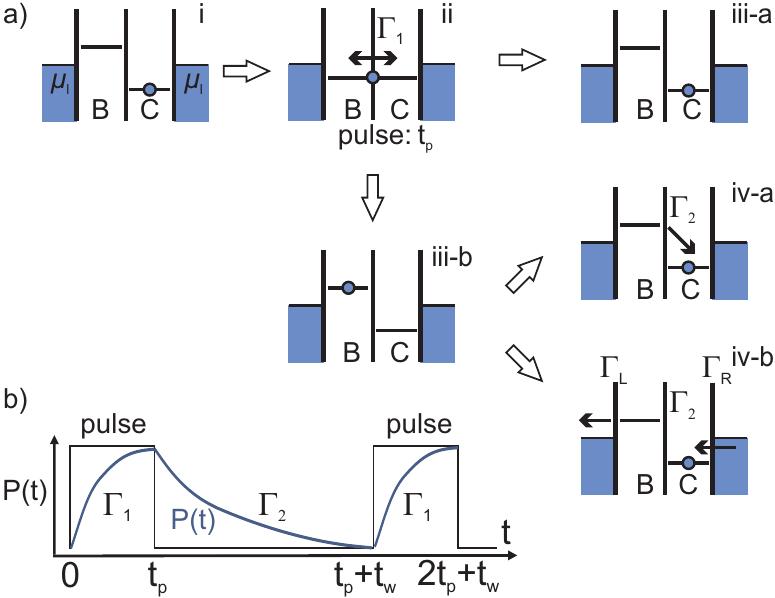}
\caption{(a) Qualitative explanation of the electron dynamics for a asymmetric pulse train. (a-i) Chemical potentials of the QDs and leads ($\mu_\mathrm l$) in the middle of the black-white line in Figs.\ \ref{fig4_a}a -- \ref{fig4_a}c. Vertical lines symbolize tunnel barriers. In the ground-state configuration 0,1, one electron occupies QD C. (a-ii) Situation while the double QD is pulsed onto (the middle of) the charge reconfiguration line. The chemical potentials of both QDs are degenerate, which allows the electron to tunnel between the QDs with rate \gi. (a-iii) After the pulse, the electron occupies either QD B (iii-b) or is back in QD C (iii-a). (a-iv) Two possible relaxation channels (combined rate \gii) from the excited state (with QD B occupied) into the ground-state. One channel (iv-a) features an interdot tunneling process in combination with energy relaxation. The other channel (iv-b) involves two resonant QD-lead tunneling processes and energy relaxation in the leads. (b) The sketch shows the steady-state probability of the (unpulsed) ground-state being unoccupied, $P(t)$, as a function of time for arbitrarily chosen \gi\ and \gii. See main text for details. The periodic pulse train is also indicated where $\tp$ is the pulse duration and $\tw$ is the waiting time between pulses.}
\label{fig4}
\end{figure}

In order to explore the origin of this nonequilibrium feature, we will now discuss the pulse sequence step by step. The sketches in Fig.\ \ref{fig4}a
indicate the chemical potentials of the two QDs and its leads. In Figs.\ \ref{fig4_a}a -- \ref{fig4_a}c the ground-state configuration at the black-white lines is 0,1 with one electron charging QD C. This situation is indicated in sketch i of Fig.\ \ref{fig4}a. The short rise time of our pulses guarantee nonadiabatic transitions. Hence, at the very beginning of the square pulse, the electron is still localized in the same QD as right before the voltage change. If pulsed onto the reconfiguration line, the configurations 0,1 and 1,0 are degenerate, and thus, during the pulse, the electron tunnels with an interdot tunneling rate \gi\ between the two QDs (sketch ii in Fig.\ \ref{fig4}a). Immediately after the pulse, the chemical potentials again correspond to the situation in sketch i. However, the new charge configuration depends on the pulse duration \tp. Here we consider the incoherent limit where \tp\ is long compared to the charge coherence time. In this limit and when $\tp\gi\gg 1$, we expect the two configurations 0,1 and 1,0 to occur with equal probability (sketches iii-a and iii-b). However, for $\tp\gi<1$, interdot tunneling is unlikely and the probability that the double QD stays in its ground-state configuration 0,1 is enhanced.

In the case that the double QD occupies the excited configuration 1,0, it tends to relax towards its ground-state configuration 0,1 during the waiting time \tw. Two possible relaxation channels 1,0 $\rightarrow$ 0,1 are indicated in sketches iv-a and iv-b in Fig.\ \ref{fig4}a. The first one includes interdot tunneling combined with energy relaxation of the electron, e.\,g.\ phonon-assisted tunneling (sketch iv-a). The other channel involves two resonant QD-lead tunneling processes 1,0 $\rightarrow$ 0,0 with rate $\gl$ followed by 0,0 $\rightarrow$ 0,1 with rate $\gr$ (sketch iv-b). For simplicity we summarize both relaxation channels 1,0 $\rightarrow$ 0,1 in the combined charge transfer rate \gii\ between two pulses --- compared to the charge transfer rate \gi\ during a pulse.

Charge transfer between QDs usually requires energy or charge exchange with the environment. In our case this results in a relatively small transfer rate \gii. The transfer rate is much larger in the case of rapid resonant tunneling between the two QDs, which occurs if the configurations 0,1 and 1,0 are degenerate. Hence, pulses into this regime, namely onto the charge reconfiguration line or its linear extensions, yield $\gi\gg\gii$.

Fig.\ \ref{fig4}b sketches the evolution of the probability $P(t)$ for the (unpulsed) ground-state configuration to be unoccupied in steady state, i.\,e.\ after many iterations of pulses. Our charge detector measures the time-average $\overline P$ of this signal. The combination of $\tp \ll \tw$ and $\gi\gg\gii$ together with $\gii^{-1}>\tw$ results in a nonequilibrium mean occupation appearing as a black-white line in Figs.\ \ref{fig4_a}a -- \ref{fig4_a}c.

\section{Qualitative discussion of different coupling regimes}

In the case of Figs.\ \ref{fig4_a}a -- \ref{fig4_a}c the pulse-induced black-white lines have exactly the same length as the charge reconfiguration lines. Here a non-equilibrium configuration apparently requires pulses directly onto the charge reconfiguration line (and not its extensions). This can be understood in terms of very asymmetric QD-lead tunneling rates, namely a much larger tunneling rate between QD B and the left lead ($\gl$) compared to QD C and the right lead $\gr\ll\gl$. During a pulse, from 0,1 onto the extension of the reconfiguration line in region 1,1, a second electron might enter the double QD from the left lead, if $\gl\tp>1$.  After the pulse, the transition 1,1 $\rightarrow$ 0,1 brings the double QD back to its ground-state configuration on a time scale $\gl^{-1}<\tp\ll\tw$. Hence, the double QD spends most of its time in the configuration 0,1 which explains why the pulse-induced black-white line does not continue beyond the length of the charge reconfiguration line. On the other hand, during a pulse directly onto the reconfiguration line no charge is exchanged with the leads (sketch ii in Fig.\ \ref{fig4}a). If, after the pulse, the configuration 1,0 is occupied, the electron will very quickly escape from QD B to the left lead (sketch iv-b in Fig.\ \ref{fig4}a). However, refilling of QD C is much slower as $\gr\ll\gl$ and tunneling of an electron from the left lead into QD C is blocked. If, in addition, $\gr\tw<1$ we expect to find that the nonequilibrium configuration 0,0 is mostly occupied within the black-white line. This is indeed the case (as will be discussed in section \ref{section8}).

\begin{figure}[th]
\includegraphics{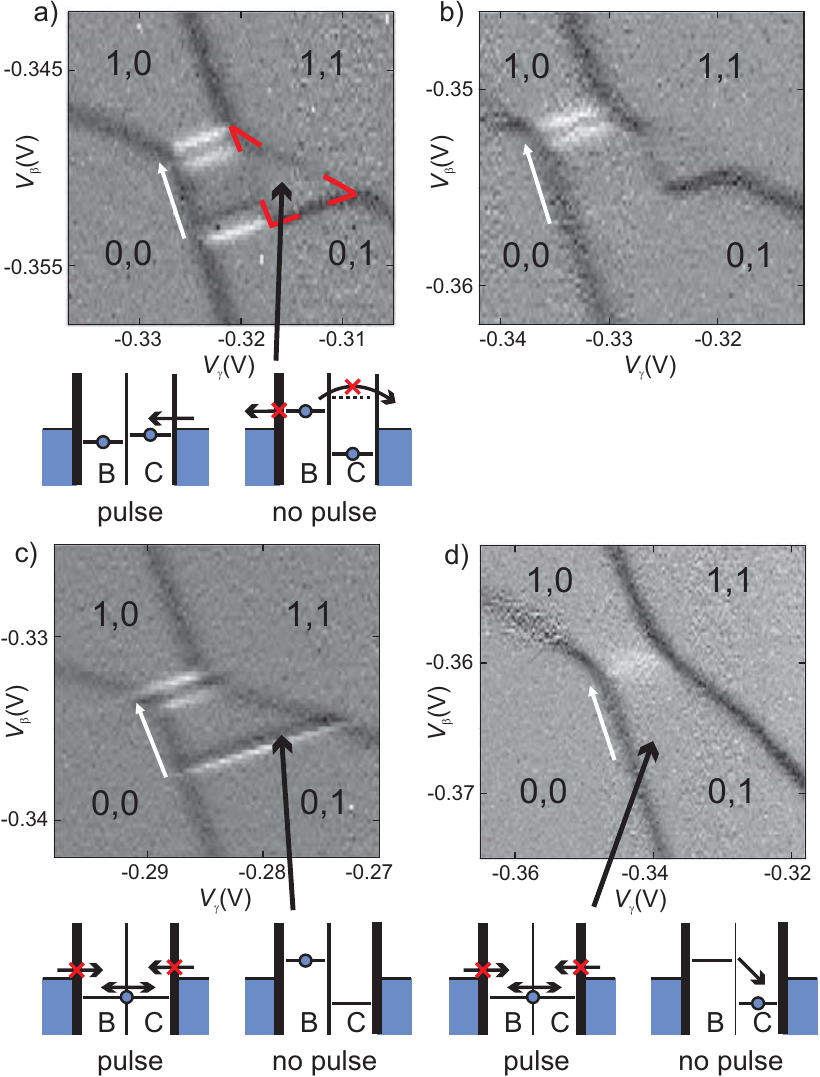}
\caption{Charge stability diagrams of a double QD for various QD-lead and interdot tunnel couplings (device and measurements as for Figs.\ \ref{fig2} and \ref{fig4_a}). Sketches: relevant charge transition processes (arrows) during pulses and between pulses. Crossed out arrows: energetically favored transitions which are very slow and, hence, result in nonequilibrium occupations. Vertical lines depict tunnel barriers (smaller width equals larger tunnel coupling). (a-b) Asymmetric QD-lead tunnel couplings as indicated in the corresponding sketches. In a) $\tp=0.5\,$ns and $\tw=10\,$ns. In b) comparable parameters except larger interdot tunnel coupling and $\tw=20\,$ns. Within the triangle marked in a) the transition 0,1 $\rightarrow$ 1,1 might happen during pulses, while the opposite transition 1,1 $\rightarrow$ 0,1 is unlikely to happen between pulses. (c) Weak QD-lead tunnel couplings on both sides suppress the transition 0,1 $\rightarrow$ 1,1 ($\tp=0.5\,$ns and $\tw=25\,$ns). (d) Weak QD-lead tunnel couplings on both sides combined with a very strong interdot tunnel coupling ($\tp=0.5\,$ns and $\tw=20\,$ns).}
\label{fig5}
\end{figure}

Figures \ref{fig5}a and \ref{fig5}b
show transconductance measurements for the same double QD as above --- with very similar periodic pulse trains applied ($\tp=0.5\,$ns and $\tw=10\,$ns for Fig.\ \ref{fig5}a) --- but a set of slightly different gate voltages. Now the tunnel barrier between QD B and the left lead is tuned to be large compared to the barrier between QD C and the right lead $\gl\ll\gr$ (opposite tendency compared to the data in Fig.\ \ref{fig4_a}). In Fig.\ \ref{fig5}a the pulse-induced black-white line is now extended by a dark line which merges into the (also dark) charging line 0,1 $\leftrightarrow$ 1,1 of QD B. Above this new line the charging line of QD B is weakened. The data suggest that within the triangle, marked in Fig.\ \ref{fig5}a by red corners, mostly the configuration 1,1 is occupied even though the ground-state there corresponds to 0,1. The sketch below Fig.\ \ref{fig5}a explains this observation. The pulses starting from within the triangle bring the double QD into the region of the stability diagram with ground-state configuration 1,1. Even for $\Gamma^{-1}_\mathrm L, \Gamma^{-1}_\mathrm R \ll\tp$ eventually the transition 0,1 $\rightarrow$ 1,1 occurs. The configuration 1,1 is not the ground-state between the pulses but still quite stable because the left tunnel barrier is large and the right QD is in Coulomb blockade. The observation of the discussed feature suggests $\Gamma_\mathrm R \tp > \Gamma_\mathrm L \tw$.

For Fig.\ \ref{fig5}b the waiting time $\tw=20\,$ns is doubled compared to Fig.\ \ref{fig5}a and the interdot coupling is increased, while all other parameters are left unchanged. On the one hand, the black-white line completely disappeared in Fig.\ \ref{fig5}b, which suggests $\gii\gg\tw^{-1}$ so that the ground-state is reoccupied quickly between two pulses. On the other hand, the just explained triangle remains present, because once the double QD is charged by two electrons the nonequilibrium state is again rather stable --- despite of the increased interdot coupling.

In Fig.\ \ref{fig5}c ($\tp=0.5\,$ns, $\tw=25\,$ns) the parameters are similar as for Figs.\ \ref{fig4_a}a -- \ref{fig4_a}c except that both tunnel barriers to the leads are now almost closed ($\gl\sim\gr\ll\tw^{-1}$). In this regime, charge exchange with the leads is absent and only the configurations 0,1 and 1,0 contribute to the nonequilibrium signal. Now, the black-white line is no longer a copy of the charge reconfiguration line but extends throughout the region of ground-state configuration 0,1.

Figure \ref{fig5}d shows the extreme case of almost closed barriers to the leads (as for Fig.\ \ref{fig5}c) but, in addition, very strong interdot tunnel coupling. Here, no effects of the applied pulse train ($\tp=0.5\,$ns and $\tw=25\,$ns) are visible due to an extreme broadening of the transition 1,0 $\leftrightarrow$ 0,1.

\section{Rate equation model}

\begin{figure}[th]
\includegraphics{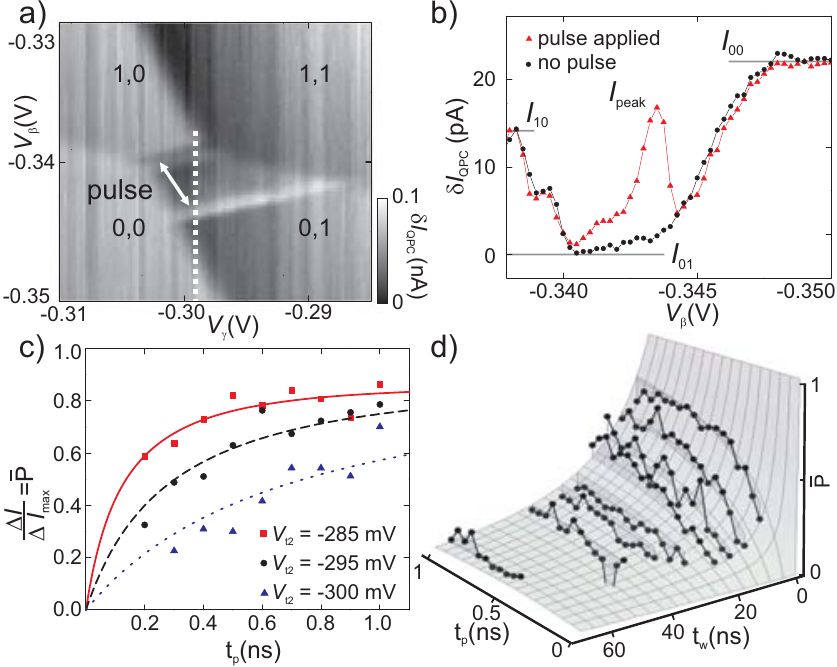}
\caption{(a) Charge stability diagram of the same double QD as in Figs.\ \ref{fig4_a}a -- \ref{fig4_a}c but for different QD-lead and interdot tunnel couplings. Plotted is the dc current $\delta I_\mathrm{QPC}$ after subtraction of a linear background. A pulse train with $\tp=0.6\,$ns and $\tw=25\,$ns is applied. (b) $\delta I_\mathrm{QPC}$ along the cut through the stability diagram indicated in Fig.\ \ref{fig6}a as a white dotted line (red triangles). The (black) filled circles show the measurement without pulsing. $I_{10}$, $I_{01}$ and $I_{00}$ are the current values expected for the charge configurations 1,0 and 0,1 and 0,0 respectively. $I_\mathrm{peak}$ marks the pulse-induced maximum of the detector current. (c) Relative detector current $\Delta I / \Delta I_\mathrm{max}$ (where $\Delta I= I_\mathrm{peak}-I_{01}$ and $\Delta I_\mathrm{max}=I_{00}-I_{01}$) as a function of \tp\ at $\tw=25\,$ns for three different values of the gate voltage $V_{t2}$ (as indicated). Solid lines are model curves from Eq.\ \ref{equ5} (see main text) (d) Filled circles are measured values of $\Delta I / \Delta I_\mathrm{max}$ as a function of \tp\ and \tw. The meshed surface is a model curve from Eq.\ \ref{equ5} for the fit parameters $\gi^{-1}=0.74\,$ns and $\gii^{-1}=12.9\,$ns. Here we assume $\Delta I / \Delta I_\mathrm{max}=\overline P$ (details in main text).}
\label{fig6}
\end{figure}

In the following we develop a quantitative rate equation model for the steady-state solution of the pulse cycle shown in Fig.\ \ref{fig4}b. As an example we focus on the regime where the tunnel couplings to both leads are weak. Figure \ref{fig5}c shows the stability diagram for such a situation in transconductance $\dd I_\mathrm{QPC}/\dd V_{\beta}$ while Fig.\ \ref{fig6}a
plots the dc detector current $\delta I_\mathrm{QPC}$ corrected by a linear offset. Instead of the black-white line observed in transconductance, the absolute current maximum is now seen as a bright line.

For simplicity, we consider purely exponential relaxation processes. As above, interdot transitions 1,0 $\leftrightarrow$ 0,1 are described by the rates \gi\ during pulses and \gii\ between pulses. The validity of this approximation needs to be verified later, if, for example, the intermediate configuration 0,0 is involved (e.\,g.\ as for Figs.\ \ref{fig4_a}a -- \ref{fig4_a}c). To allow a direct comparison with experimental data we determine the probability, $P(t)$, that the unpulsed ground-state configuration 0,1 is \emph{not} occupied (either 1,0 or 0,0 is occupied). Immediately \emph{before} a pulse is applied (at t=0) and at the end of a pulse (at $t=\tp$) $P(t)$ can be expressed by
\begin{equation} \label{equ1}
 \begin{array}{ll}
 P(0)&=P(\tp)\mathrm \e^{-\tw\gii}\,,\\\\
 P(\tp)&=1-\left[1-P(0)\right]\mathrm \e^{-\tp\gi}\,.
 \end{array}
\end{equation}
After many repetitions, a steady state is reached, and $P(0)=P(\tp+\tw)$ as indicated in Fig.\ \ref{fig4}b. Solving for $P(0)$ and $P(\tp)$ using Eq\.(\ref{equ1}) leads to
\begin{equation}
 \begin{array}{ll}
P(0)&={1-\mathrm \e^{\tp\gi}\over 1-\mathrm \e^{{\tw\gii}+{\tp\gi}}}\,,\\\\
P(\tp)&=1-{1-\mathrm \e^{\tw\gii}\over 1-\mathrm \e^{{\tw\gii}+{\tp\gi}}}\,.
 \end{array}
\nonumber\end{equation}
Finally, integration over one pulse cycle in the steady-state limit results in the average probability that the unpulsed ground-state is unoccupied
\begin{equation}\label{equ5}
 \begin{array}{ll}
\overline P &=\frac{1}{\tp+\tw}\left[\int\limits_0^{\tp}({1-\left[1-P(0)\right]\e^{-t\gi})\dd t}+\int\limits_0^{\tw}{P(\tp)\e^{-t\gii}\dd t}\right]\\\\
&=\frac{1}{\tp+\tw}\left(\tp+\left[\frac{1}{\gi}-\frac{1}{\gii}\right]\bigl[P(0)-P(\tp)\bigr]\right)\,.
 \end{array}
\end{equation}

\vskip 7ex

\section{Quantitative analysis and sample characterization}\label{section8}

Figure \ref{fig6}b plots $\delta I_\mathrm{QPC}$ measured along the white dotted line in Fig.\ \ref{fig6}a for two cases, namely, with (red triangles) or without pulsing (black circles). The average values of $I_\mathrm{QPC}$ expected for the configurations 1,0 ($I_{10}$), 0,1 ($I_{01}$) and 0,0 ($I_{00}$) are indicated. At the left end of the plot, $I_\mathrm{QPC}\simeq I_{10}$, which indicates occupation of the ground-state configuration 1,0. Next $I_\mathrm{QPC}$ decreases to $I_{0,1}$ in two steps as the (split) charge reconfiguration line is crossed (see Figs.\ \ref{fig4_a}a -- c). For the case without pulsing, $I_\mathrm{QPC}$ stays approximately constant at $I_{0,1}$ before it increases to $I_{0,0}$ across the charging line between 0,1 and 0,0.

In contrast, the curve produced while pulsing contains an additional maximum labeled $I_\mathrm{peak}$ within the region of the ground-state configuration 0,1. This maximum corresponds to the bright line in Fig.\ \ref{fig6}a. Its amplitude provides not only information about the average probability $\overline P$, but also about the relevance of the two possible relaxation channels depicted in the sketches iv-a and iv-b in Fig.\ \ref{fig4}a. For the data presented here, we observe that $I_\mathrm{peak}$ can exceed $I_{1,0}$ and even takes values up to $I_{0,0}$. This behavior contradicts the expectation for a combination of incoherent resonant tunneling (sketch ii in Fig.\ \ref{fig4}a) and direct energy relaxation via interdot tunneling (sketch iv-a), because the overall occupation of one electron in the double QD is preserved in these processes. In the case of incoherent interdot tunneling, $I_\mathrm{peak}$ would be limited by $0.5\times I_{10}$. Hence, our system favors the other relaxation channel 1,0 $\rightarrow$ 0,0 $\rightarrow$ 0,1 (sketch iv-b in Fig.\ \ref{fig4}a) which involves charge exchange with the leads. Moreover, $I_\mathrm{peak} \gtrsim I_{1,0}$ implies that in this case the intermediate configuration 0,0 is mostly occupied, which requires that the tunneling process 1,0 $\rightarrow$ 0,0 is faster than the subsequent transition to the ground-state 0,0 $\rightarrow$ 0,1. In this limit the two-step tunneling process can be approximated by a single exponential decay rate \gii\ and our model expressed in Eq.\ (\ref{equ5}) can indeed be applied. However, it is possible to tune the double QD into different regimes (compare Fig.\ \ref{fig5}), some of which would require modification of the rate equation model.

Figure \ref{fig6}c presents the relative current change $\Delta I / \Delta I_\mathrm{max}$, where $\Delta I= I_\mathrm{peak}-I_{01}$ and $\Delta I_\mathrm{max}=I_{00}-I_{01}$, as a function of $\tp$ for fixed $\tw=25\,$ns. The three different symbols correspond to three different voltages applied to gate $t2$ which controls the interdot tunneling rate \gi. As expected $\Delta I / \Delta I_\mathrm{max}$ grows faster for a smaller interdot tunnel barrier $t2$ (less negative $V_\mathrm{t2}$) and a larger tunneling rate \gi. As long as the occupation time of configuration 1,0 is very short compared to that of 0,0 our model allows the direct interpretation $\Delta I / \Delta I_\mathrm{max} \equiv \overline P$ and, therefore, a quantitative comparison with Eq.\ (\ref{equ5}). The lines are the according model curves for $\gii^{-1}=75\,$ns and $\gi^{-1}=0.39\,$, 0.92\, and 2.33\,ns for  $V_\mathrm{t2}=-285\,$, $-295\,$ and $-300\,$mV.

A more complete comparison between similar data and our rate equation model is presented in Fig.\ \ref{fig6}d, a plot of $\overline P$ as a function of \tp\ and \tw. The measured data (closed circles) are overlayed with a two-dimensional model curve representing Eq.\ (\ref{equ5}) with the fit parameters $\gi^{-1}=0.74\,$ns, $\gii^{-1}=12.9\,$ns \cite{comment-3}. The overall smooth increase in $\overline P$ with increasing \tp\ and decreasing \tw\ indicates incoherent tunneling between the two QDs in these measurements.

\section{Summary}

In conclusion we have demonstrated how charge spectroscopy is capable of resolving single electron charging events in complex nanostructures at time scales below 1\,ns. These experiments require a large bandwidth and it is crucial that the sample holder and other rf components are well designed to avoid reflections and cross-talk between control gates. In our measurements we have combined periodic trains of ultra-short voltage pulses on specific gates with a continuous (dc) charge detection using a QPC. A rate equation model can be used to quantitatively analyze the data and to separate tunneling rates from energy relaxation times. Different relaxation channels in the system can be identified and controlled. This pulsed-gate technique, demonstrated here for a full characterization of a double QD, can be used for coherent control of charge qubits or coherent transport by adiabatic passage \cite{Greentree2004,Cole2008}. Our approach promises scalability towards an array of qubits in which each qubit is separately addressable while mutual qubit entanglement can be fully controlled.

\section{Acknowledgements}

We thank Q.\ Unterreithmeier for helpful discussions. Financial support by the German Science Foundation via SFB 631 and SFB 689, the German Israel program DIP, the German Excellence Initiative via the "Nanosystems Initiative Munich (NIM)", and LMUinnovativ (FuNS) is gratefully acknowledged.


\begin{thebibliography}{27}
\bibitem{comment-1}{The sample holder is based on a Rogers TMM 10 substrate with a thickness of 0.5\,mm and $17\,\mu$m copper cladding.}
\bibitem{comment-2}{The pulses are generated with a delay of 370 ps to compensate for slightly different cable lengths inside the cryostat.}
\bibitem{comment-3}{Here we obtain also the normalization to $\Delta I / \Delta I_\mathrm{max}$ from the fit.}
\end{thebibliography}

\end{document}